\documentclass[sigconf,screen]{acmart}

\usepackage{enumitem}
\usepackage{amsfonts}
\usepackage{algorithmic}
\usepackage[caption=false,font=normalsize,labelfont=sf,textfont=sf]{subfig}
\usepackage{textcomp}
\usepackage{stfloats}
\usepackage{url}
\usepackage{verbatim}
\usepackage{multirow}
\usepackage{soul}
\usepackage{fontawesome}
\usepackage{makecell}
\usepackage{siunitx}
\usepackage{colortbl}

\AtBeginDocument{
  \providecommand\BibTeX{{
    \normalfont B\kern-0.5em{\scshape i\kern-0.25em b}\kern-0.8em\TeX}}}

\newcommand{\Comment}[1]{}

\NewDocumentCommand{\framecolorbox}{oommm}
 {
  \IfValueTF{#1}
   {\IfValueTF{#2}
    {\fcolorbox{#3}{#4}{\makebox[#1][#2]{#5}}}
    {\fcolorbox{#3}{#4}{\makebox[#1]{#5}}}%
   }
   {\fcolorbox{#3}{#4}{#5}}%
 }

\copyrightyear{2026}
\acmYear{2026}
\setcopyright{cc}
\setcctype{by}
\acmConference[ICSE-Companion '26]{2026 IEEE/ACM 48th International Conference on Software Engineering}{April 12--18, 2026}{Rio de Janeiro, Brazil}
\acmBooktitle{2026 IEEE/ACM 48th International Conference on Software Engineering (ICSE-Companion '26), April 12--18, 2026, Rio de Janeiro, Brazil}
\acmPrice{}
\acmDOI{10.1145/3774748.3787629}
\acmISBN{979-8-4007-2296-7/2026/04}

\begin{document}

\title{Advancing Language Models for Code-related Tasks}

\author{Zhao Tian}
\orcid{0000-0002-9316-7250}
\affiliation{
  \institution{School of Computer Software, Tianjin University}
  \city{Tianjin}
  \country{China}
}
\email{tianzhao@tju.edu.cn}

\begin{abstract}
Recent advances in language models (LMs) have driven significant progress in various software engineering tasks. 
However, existing LMs still struggle with complex programming scenarios due to limitations in data quality, model architecture, and reasoning capability. 
This research systematically addresses these challenges through three complementary directions: 
(1) improving code data quality with a code difference-guided adversarial augmentation technique (CODA) and a code denoising technique (CodeDenoise); 
(2) enhancing model architecture via syntax-guided code LMs (LEAM and LEAM++); 
and (3) advancing model reasoning with a prompting technique ($\mu$FiX) and an agent-based technique (Specine). 
These techniques aim to promote the practical adoption of LMs in software development and further advance intelligent software engineering.
\end{abstract}

\ccsdesc[500]{Software and its engineering~Software development techniques}
\ccsdesc[300]{Computing methodologies~Natural language processing}
\ccsdesc[300]{Computing methodologies~Neural networks}

\keywords{Code-related Task, Language Model}

\maketitle

\section{Introduction}
\label{sec:introduction}

In recent years, the rapid advancement of language models (LMs) has led to the emergence of code LMs trained on large-scale programming corpora, attracting increasing attention in the software engineering community~\cite{tian2024large,tian2025ensemble}.
By learning from massive open-source code data, these LMs exhibit strong capabilities in code understanding and generation, enabling their application to various software engineering tasks, including code generation~\cite{gao2025trae}, automated testing~\cite{wang2024software}, and vulnerability detection~\cite{steenhoek2023empirical}. 
For instance, in code generation, code LMs can produce implementations from natural language descriptions, substantially reducing repetitive work and improving development efficiency. 

Despite these promising progress, code LMs still face substantial challenges in handling complex software engineering tasks. 
For example, evaluations of state-of-the-art models (e.g., Claude 3.7) in a real-world tasks show that only 11.6\% of programming problems are successfully resolved, revealing their limited practical effectiveness~\cite{zan2025multiswebench}. 
Such performance bottlenecks hinder the widespread adoption of code LMs and may compromise software quality.
Consequently, systematically enhancing the effectiveness of code LMs for diverse code-related tasks has emerged as a critical research priority for both academia and industry.

To improve the model performance in software engineering tasks, this research focuses on three key directions:
(1) \textit{Code data quality}: As the foundation of code LMs, code data quality directly affects model performance~\cite{kuang2025effectiveness}. 
Issues such as imbalanced data distribution and inconsistency in syntax and semantics substantially limit further performance improvement.
(2) \textit{Model architecture}: Many studies design advanced architectures to enhance code understanding and generation~\cite{feng2020codebert}. 
However, most LMs still represent programs as flat token sequences, limiting their ability to capture program features and ensure code correctness.
(3) \textit{Model reasoning}: Recent progress in prompt engineering and agent-based system have improved the reasoning ability of code LMs~\cite{nashid2023retrieval}. 
However, they often overlook misunderstanding fixing and requirement alignment, posing substantial challenges when addressing complex tasks.

\section{Methodology}
\label{sec:methodology}
\begin{figure}[t!]
    \centering
	\includegraphics[width=1.0\linewidth]{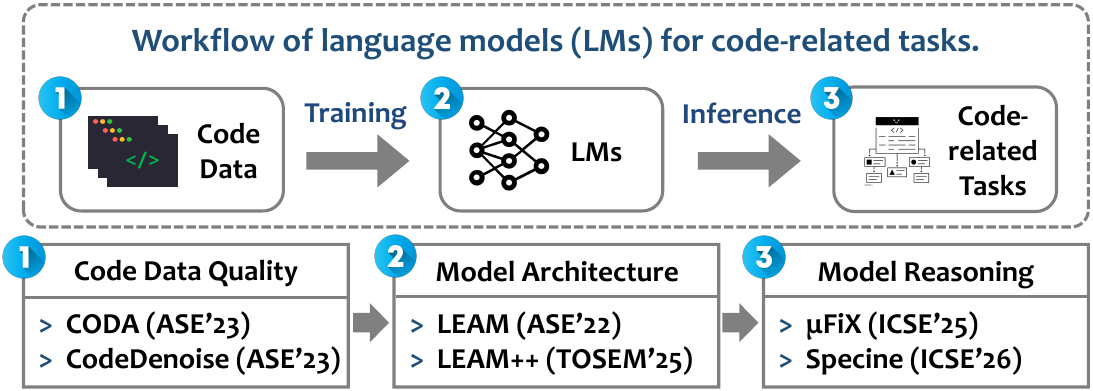}	
    \vspace{-4mm}
    \caption{An overview of the research thesis}
    \label{fig:overview}
    \vspace{-4mm}
\end{figure}

As shown in Figure~\ref{fig:overview}, this research aims to systematically address these limitations across the three complementary directions for fostering the practical adoption of LMs in software development.

\vspace{-2mm}
\subsection{Improving Code Data Quality}
\textbf{\textit{\underline{Challenges.}}}
This research addresses two key challenges in existing code data:
(1) \textit{Imbalanced data distribution}:
Current code datasets often exhibit distributional bias, which limits the generalization ability of code LMs.
To mitigate this issue, prior studies~\cite{tian2025codefense,zhang2022towards} have employed adversarial code augmentation strategies that apply equivalent transformations (e.g., identifier renaming or dead-code insertion) to enrich data diversity.
However, existing methods typically rely on greedy search guided by model prediction changes on transformed inputs, which can easily fall into local optima and thus reduce overall effectiveness.
(2) \textit{Inconsistency in syntax and semantics}:
Code snippets often contain inconsistencies between their syntactic implementation and semantic meaning.
Although such code noise does not affect functionality, they can disrupt the model's program comprehension and reduce prediction accuracy.
There remains a lack of a systematic code denoising approach.

\noindent
\textbf{\textit{\underline{Approaches.}}}
To address the above two challenges, this research designs two complementary techniques: a code difference-guided adversarial augmentation technique (CODA~\cite{tian2023code}) and a code denoising technique (CodeDenoise~\cite{tian2023fly}), both aimed at improving the overall quality of code data.
Specifically, CODA leverages code differences between the target input (i.e., a given code snippet as the model input) and reference inputs (i.e., the inputs that have small code differences but different prediction results with the target input) to guide adversarial augmentation. 
CodeDenoise aims to mitigate inconsistencies between code syntax and semantics, which consists of estimating a mispredicted input, localizing noise in the input code, and cleansing noise to correct the misprediction.

\noindent
\textbf{\textit{\underline{Results.}}}
Experimental results demonstrate the effectiveness of both CODA and CodeDenoise.
Specifically, CODA surpasses state-of-the-art baselines (ALERT~\cite{yang2022natural} and CARROT~\cite{zhang2022towards}) by 28.86\% and 24.04\% on average in terms of model robustness.
Meanwhile, CodeDenoise successfully denoises 21.91\% of mispredicted code samples and improves model accuracy by an average of 2.04\%, significantly outperforming the commonly used fine-tuning strategy.

\vspace{-2mm}
\subsection{Improving Model Architecture}
\textbf{\textit{\underline{Challenges.}}}
Most existing code LMs directly adopt the architectures of general language models, treating programs as flat token sequences~\cite{roziere2023code,guo2024deepseek}.
However, representing code merely as a token sequence neglects rich structural and semantic information, thereby impairing the overall model performance.
Hence, generating code by predicting tokens sequentially leads to an enormous search space, making it difficult to produce accurate implementations.
Moreover, such LMs often fail to guarantee syntactically correct programs.

\noindent
\textbf{\textit{\underline{Approaches.}}}
To address these limitations, this research proposes syntax-guided code LMs (LEAM~\cite{tian2022learning} and LEAM++~\cite{tianleam++}).
Unlike general LMs, LEAM and LEAM++ represent code as abstract syntax trees (ASTs) to integrate both structural and semantic information, thereby enhancing prediction performance.
To ensure the syntactic correctness of generated code, they adopt a syntax-guided encoder–decoder architecture that predicts appropriate grammar rules for each unexpanded nonterminal node in the partial AST.

\noindent
\textbf{\textit{\underline{Results.}}}
To evaluate the effectiveness of LEAM and LEAM++, we conducted extensive experiments on the mutation code generation task.
The results show that both techniques achieve 100\% syntactic correctness and generate mutation code that more accurately reflects real faults compared to those constructed by existing techniques (i.e., Major~\cite{just2014major}, PIT~\cite{coles2016pit}, and DeepMutation~\cite{tufano2019learning}).

\vspace{-2mm}
\subsection{Improving Mode Reasoning}
\textbf{\textit{\underline{Challenges.}}}
Despite recent advances in prompt engineering and agent-based systems, existing approaches still face significant limitations.
First, thought-eliciting prompting techniques~\cite{li2025structured,jiang2024self} often struggle to produce correct understanding for complex programming tasks, leading to suboptimal performance.
Meanwhile, feedback-based prompting techniques~\cite{olausson2024selfrepair,chen2024teaching} struggle to refine generation when the generated code deviates substantially from the ground truth, as test feedback provides only coarse-grained information.
Second, agent-based techniques~\cite{hong2023metagpt,dong2024selfcollaboration} overlook the critical issue of requirement perception, resulting in persistent misalignment issues. 
Given that accurate perception serves as the foundation of the LLM-based code generation paradigm~\cite{pohl1996requirements,zave1997four,macaulay2012requirements}, ensuring requirement alignment remains a critical challenge.

\noindent
\textbf{\textit{\underline{Approaches.}}}
To address the above two challenges, this research proposes two model reasoning techniques: a novel prompting technique ($\mu$FiX~\cite{tian2025fixing}) and an agent-based technique (Specine~\cite{tian2025aligning}).
Specifically, $\mu$FiX is the first to combine thought-eliciting and feedback-based prompting, leveraging test case analysis to enhance specification understanding and further fix misunderstandings for better code generation.
Specine focuses on identifying misaligned input requirements, lifting LLM-perceived requirements, and aligning them to enhance the overall code generation performance.

\noindent
\textbf{\textit{\underline{Results.}}}
Experimental results demonstrate the effectiveness of both $\mu$FiX and Specine.
Specifically, $\mu$FiX outperforms the most effective prompting baseline with an average improvement of 35.62\% in terms of Pass@1.
Similarly, Specine outperforms the most effective agent-based baseline, achieving an average improvement of 29.60\% in terms of Pass@1.

\section{Related Work}
\label{sec:related}

\textbf{Code data quality.} 
CARROT~\cite{zhang2022towards} designs two semantic-preserving code transformation rules (i.e., identifier renaming and dead code insertion), and uses the hill-climbing algorithm to search for the ingredients from the entire space with the guidance of gradients and model prediction changes. 
ALERT~\cite{yang2022natural} considers the rule of identifier renaming, and uses the naturalness and model prediction changes to guide the ingredient search process.

\noindent
\textbf{Model architecture.} 
Existing code LMs (e.g., CodeLlama~\cite{roziere2023code}, Qwen-Coder~\cite{hui2024qwen2}, and DeepSeek-Coder~\cite{guo2024deepseek}) achieve excellent performance on code-related tasks. 
However, they use the same representation method as general LMs for natural language tasks~\cite{min2023recent}, treating code as token sequences, which limits their performance.

\noindent
\textbf{Model reasoning.} 
Prompting techniques (e.g., Self-planning~\cite{jiang2024self}, SCoT~\cite{li2025structured}, Self-Edit~\cite{zhang2023self}, and Self-repair~\cite{olausson2024selfrepair}) enhance model performance by designing carefully-crafted prompts that can be applied in a plug-and-play manner.
Agent-based techniques (e.g., Self-collaboration~\cite{dong2024selfcollaboration}, AgentCoder~\cite{huang2023agentcoder}, TGen~\cite{mathews2024tgen}, and FlowGen~\cite{lin2025flowgen}) improve the model performance by simulating human collaborative software development processes, where multiple agents interact to accomplish tasks (e.g., planning, implementation, and testing).

\section{Contribution Summary}
\label{sec:contribution}
The main contributions of this research are as follows:
\begin{itemize}[leftmargin=10pt]
    \item \textbf{Code Data Quality Improvement}: We propose a code difference-guided adversarial augmentation technique (CODA~\cite{tian2023code}) and a code denoising technique (CodeDenoise~\cite{tian2023fly}) to complementarily improve the quality of code data.
    
    \item \textbf{Model Architecture Improvement}: We propose novel code LMs (LEAM~\cite{tian2022learning} and LEAM++~\cite{tianleam++}) that adopt the syntax-guided encoder-decoder architecture to better capture program features and ensure syntactic correctness for code generation.

    \item \textbf{Model Reasoning Improvement}: We propose a prompting technique ($\mu$FiX~\cite{tian2025fixing}) and an agent-based technique (Specine~\cite{tian2025aligning}) to enhance the code generation performance through misunderstanding fixing and requirement alignment.
\end{itemize}

\section{Future Work}
\label{sec:future}
Although existing techniques have substantially improved LM performance in code-related tasks, they remain largely confined to low-complexity scenarios.
Consequently, their applicability to real-world, large-scale software development is still limited.
Future work will focus on advancing the capabilities of LMs to effectively support complex and practical software engineering tasks.

\begin{acks}
Zhao Tian is advised by \textbf{Professor. Junjie Chen}.
This work is supported by National Natural Science Foundation of China (Grant No. 62322208).
\end{acks}

\balance
\bibliographystyle{ACM-Reference-Format}
\bibliography{reference}

@article{tian2025codefense,
  title={CoDefense: Defending Method Against Adversarial Attacks with Multi-granularity Code Normalization},
  author={Tian, Zhao and Kuang, Shiqi and Yan, Ming and Wang, Haichi and Chen, Junjie},
  journal={Journal of Software},
  pages={1--27},
  year={2025}
}

@article{kuang2025effectiveness,
  title={On the Effectiveness of Training Data Optimization for LLM-based Code Generation: An Empirical Study},
  author={Kuang, Shiqi and Tian, Zhao and Xiao, Tao and Wang, Dong and Chen, Junjie},
  journal={arXiv preprint arXiv:2512.24570},
  year={2025}
}

@article{gao2025trae,
  title={Trae agent: An llm-based agent for software engineering with test-time scaling},
  author={Gao, Pengfei and Tian, Zhao and Meng, Xiangxin and Wang, Xinchen and Hu, Ruida and Xiao, Yuanan and Liu, Yizhou and Zhang, Zhao and Chen, Junjie and Gao, Cuiyun and others},
  journal={arXiv preprint arXiv:2507.23370},
  year={2025}
}

@inproceedings{tian2025ensemble,
  title={Agent-Based Ensemble Reasoning for Repository-Level Issue Resolution},
  author={Tian, Zhao and Gao, Pengfei and Chen, Junjie and Peng, Chao},
  booktitle={2026 IEEE/ACM 48th International Conference on Software Engineering (ICSE)},
  year={2026}
}

@article{min2023recent,
  title={Recent advances in natural language processing via large pre-trained language models: A survey},
  author={Min, Bonan and Ross, Hayley and Sulem, Elior and Veyseh, Amir Pouran Ben and Nguyen, Thien Huu and Sainz, Oscar and Agirre, Eneko and Heintz, Ilana and Roth, Dan},
  journal={ACM Computing Surveys},
  volume={56},
  number={2},
  pages={1--40},
  year={2023},
  publisher={ACM New York, NY}
}

@inproceedings{tian2025aligning,
  title={Aligning Requirement for Large Language Model's Code Generation},
  author={Tian, Zhao and Chen, Junjie},
  booktitle={2026 IEEE/ACM 48th International Conference on Software Engineering (ICSE)},
  year={2026}
}

@inproceedings{tufano2019learning,
  title={Learning how to mutate source code from bug-fixes},
  author={Tufano, Michele and Watson, Cody and Bavota, Gabriele and Di Penta, Massimiliano and White, Martin and Poshyvanyk, Denys},
  booktitle={2019 IEEE International conference on software maintenance and evolution (ICSME)},
  pages={301--312},
  year={2019},
  organization={IEEE}
}

@inproceedings{coles2016pit,
  title={Pit: a practical mutation testing tool for java},
  author={Coles, Henry and Laurent, Thomas and Henard, Christopher and Papadakis, Mike and Ventresque, Anthony},
  booktitle={Proceedings of the 25th international symposium on software testing and analysis},
  pages={449--452},
  year={2016}
}

@inproceedings{just2014major,
  title={The Major mutation framework: Efficient and scalable mutation analysis for Java},
  author={Just, Ren{\'e}},
  booktitle={Proceedings of the 2014 international symposium on software testing and analysis},
  pages={433--436},
  year={2014}
}

@article{zhang2022towards,
  title={Towards robustness of deep program processing models—detection, estimation, and enhancement},
  author={Zhang, Huangzhao and Fu, Zhiyi and Li, Ge and Ma, Lei and Zhao, Zhehao and Yang, Hua’an and Sun, Yizhe and Liu, Yang and Jin, Zhi},
  journal={ACM Transactions on Software Engineering and Methodology (TOSEM)},
  volume={31},
  number={3},
  pages={1--40},
  year={2022},
  publisher={ACM New York, NY}
}

@inproceedings{yang2022natural,
  title={Natural attack for pre-trained models of code},
  author={Yang, Zhou and Shi, Jieke and He, Junda and Lo, David},
  booktitle={Proceedings of the 44th International Conference on Software Engineering},
  pages={1482--1493},
  year={2022}
}

@inproceedings{feng2020codebert,
  title={CodeBERT: A Pre-Trained Model for Programming and Natural Languages},
  author={Feng, Zhangyin and Guo, Daya and Tang, Duyu and Duan, Nan and Feng, Xiaocheng and Gong, Ming and Shou, Linjun and Qin, Bing and Liu, Ting and Jiang, Daxin and others},
  booktitle={Findings of the Association for Computational Linguistics: EMNLP 2020},
  pages={1536--1547},
  year={2020}
}

@misc{zan2025multiswebench,
      title={Multi-SWE-bench: A Multilingual Benchmark for Issue Resolving}, 
      author={Daoguang Zan and Zhirong Huang and Wei Liu and Hanwu Chen and Linhao Zhang and Shulin Xin and Lu Chen and Qi Liu and Xiaojian Zhong and Aoyan Li and Siyao Liu and Yongsheng Xiao and Liangqiang Chen and Yuyu Zhang and Jing Su and Tianyu Liu and Rui Long and Kai Shen and Liang Xiang},
      year={2025},
      eprint={2504.02605},
      archivePrefix={arXiv},
      primaryClass={cs.SE},
      url={https://arxiv.org/abs/2504.02605}, 
}

@inproceedings{steenhoek2023empirical,
  title={An empirical study of deep learning models for vulnerability detection},
  author={Steenhoek, Benjamin and Rahman, Md Mahbubur and Jiles, Richard and Le, Wei},
  booktitle={2023 IEEE/ACM 45th International Conference on Software Engineering (ICSE)},
  pages={2237--2248},
  year={2023},
  organization={IEEE}
}

@article{wang2024software,
  title={Software testing with large language models: Survey, landscape, and vision},
  author={Wang, Junjie and Huang, Yuchao and Chen, Chunyang and Liu, Zhe and Wang, Song and Wang, Qing},
  journal={IEEE Transactions on Software Engineering},
  volume={50},
  number={4},
  pages={911--936},
  year={2024},
  publisher={IEEE}
}

@article{tianleam++,
  title={LEAM++: Learning for Selective Mutation Fault Construction},
  author={Tian, Zhao and Chen, Junjie and Wang, Dong and Zhu, Qihao and Fan, Xingyu and Zhang, Lingming},
  journal={ACM Transactions on Software Engineering and Methodology},
  publisher={ACM New York, NY}
}

@article{zave1997four,
  title={Four dark corners of requirements engineering},
  author={Zave, Pamela and Jackson, Michael},
  journal={ACM transactions on Software Engineering and Methodology (TOSEM)},
  volume={6},
  number={1},
  pages={1--30},
  year={1997},
  publisher={ACM New York, NY, USA}
}

@inproceedings{tian2024large,
  title={Large language models for equivalent mutant detection: How far are we?},
  author={Tian, Zhao and Shu, Honglin and Wang, Dong and Cao, Xuejie and Kamei, Yasutaka and Chen, Junjie},
  booktitle={Proceedings of the 33rd ACM SIGSOFT International Symposium on Software Testing and Analysis},
  pages={1733--1745},
  year={2024}
}

@article{hui2024qwen2,
  title={Qwen2. 5-coder technical report},
  author={Hui, Binyuan and Yang, Jian and Cui, Zeyu and Yang, Jiaxi and Liu, Dayiheng and Zhang, Lei and Liu, Tianyu and Zhang, Jiajun and Yu, Bowen and Lu, Keming and others},
  journal={arXiv preprint arXiv:2409.12186},
  year={2024}
}

@article{guo2024deepseek,
  title={DeepSeek-Coder: When the Large Language Model Meets Programming--The Rise of Code Intelligence},
  author={Guo, Daya and Zhu, Qihao and Yang, Dejian and Xie, Zhenda and Dong, Kai and Zhang, Wentao and Chen, Guanting and Bi, Xiao and Wu, Yu and Li, YK and others},
  journal={arXiv preprint arXiv:2401.14196},
  year={2024}
}

@book{pohl1996requirements,
  title={Requirements engineering: An overview},
  author={Pohl, Klaus},
  year={1996},
  publisher={Citeseer}
}

@book{macaulay2012requirements,
  title={Requirements engineering},
  author={Macaulay, Linda A},
  year={2012},
  publisher={Springer Science \& Business Media}
}

@inproceedings{lin2025flowgen,
  title={SOEN-101: Code Generation by Emulating Software Process Models Using Large Language Model Agents},
  author={Lin, Feng and Kim, Dong Jae and Chen, Tse-Hsun (Peter)},
  booktitle={2025 IEEE/ACM 47th International Conference on Software Engineering (ICSE)},
  year={2025}
}

@inproceedings{mathews2024tgen,
  title={Test-Driven Development and LLM-based Code Generation},
  author={Mathews, Noble Saji and Nagappan, Meiyappan},
  booktitle={Proceedings of the 39th IEEE/ACM International Conference on Automated Software Engineering},
  pages={1583--1594},
  year={2024}
}

@article{huang2023agentcoder,
  title={Agentcoder: Multi-agent-based code generation with iterative testing and optimisation},
  author={Huang, Dong and Bu, Qingwen and Zhang, Jie M and Luck, Michael and Cui, Heming},
  journal={arXiv preprint arXiv:2312.13010},
  year={2023}
}

@article{hong2023metagpt,
  title={Metagpt: Meta programming for multi-agent collaborative framework},
  author={Hong, Sirui and Zheng, Xiawu and Chen, Jonathan and Cheng, Yuheng and Wang, Jinlin and Zhang, Ceyao and Wang, Zili and Yau, Steven Ka Shing and Lin, Zijuan and Zhou, Liyang and others},
  journal={arXiv preprint arXiv:2308.00352},
  year={2023}
}

@article{dong2024selfcollaboration,
  title={Self-collaboration code generation via chatgpt},
  author={Dong, Yihong and Jiang, Xue and Jin, Zhi and Li, Ge},
  journal={ACM Transactions on Software Engineering and Methodology},
  volume={33},
  number={7},
  pages={1--38},
  year={2024},
  publisher={ACM New York, NY}
}

@inproceedings{tian2025fixing,
  title={Fixing Large Language Models' Specification Misunderstanding for Better Code Generation},
  author={Tian, Zhao and Chen, Junjie and Zhang, Xiangyu},
  booktitle={2025 IEEE/ACM 47th International Conference on Software Engineering (ICSE)},
  year={2025}
}

@inproceedings{tian2022learning,
  title={Learning to construct better mutation faults},
  author={Tian, Zhao and Chen, Junjie and Zhu, Qihao and Yang, Junjie and Zhang, Lingming},
  booktitle={Proceedings of the 37th IEEE/ACM International Conference on Automated Software Engineering},
  pages={1--13},
  year={2022}
}

@inproceedings{tian2023fly,
  title={On-the-fly improving performance of deep code models via input denoising},
  author={Tian, Zhao and Chen, Junjie and Zhang, Xiangyu},
  booktitle={2023 38th IEEE/ACM International Conference on Automated Software Engineering (ASE)},
  pages={560--572},
  year={2023},
  organization={IEEE}
}

@inproceedings{tian2023code,
  title={Code difference guided adversarial example generation for deep code models},
  author={Tian, Zhao and Chen, Junjie and Jin, Zhi},
  booktitle={2023 38th IEEE/ACM International Conference on Automated Software Engineering (ASE)},
  pages={850--862},
  year={2023},
  organization={IEEE}
}

@article{roziere2023code,
  title={Code llama: Open foundation models for code},
  author={Roziere, Baptiste and Gehring, Jonas and Gloeckle, Fabian and Sootla, Sten and Gat, Itai and Tan, Xiaoqing Ellen and Adi, Yossi and Liu, Jingyu and Remez, Tal and Rapin, J{\'e}r{\'e}my and others},
  journal={arXiv preprint arXiv:2308.12950},
  year={2023}
}

@inproceedings{nashid2023retrieval,
  title={Retrieval-based prompt selection for code-related few-shot learning},
  author={Nashid, Noor and Sintaha, Mifta and Mesbah, Ali},
  booktitle={Proceedings of the 45th International Conference on Software Engineering (ICSE'23)},
  year={2023}
}

@inproceedings{chen2024teaching,
    title={Teaching Large Language Models to Self-Debug},
    author={Xinyun Chen and Maxwell Lin and Nathanael Sch{\"a}rli and Denny Zhou},
    booktitle={The Twelfth International Conference on Learning Representations},
    year={2024},
}

@article{zhang2023self,
  title={Self-Edit: Fault-Aware Code Editor for Code Generation},
  author={Zhang, Kechi and Li, Zhuo and Li, Jia and Li, Ge and Jin, Zhi},
  journal={arXiv preprint arXiv:2305.04087},
  year={2023}
}

@inproceedings{olausson2024selfrepair,
    title={Is Self-Repair a Silver Bullet for Code Generation?},
    author={Theo X. Olausson and Jeevana Priya Inala and Chenglong Wang and Jianfeng Gao and Armando Solar-Lezama},
    booktitle={The Twelfth International Conference on Learning Representations},
    year={2024},
}

@article{jiang2024self,
  title={Self-planning code generation with large language models},
  author={Jiang, Xue and Dong, Yihong and Wang, Lecheng and Fang, Zheng and Shang, Qiwei and Li, Ge and Jin, Zhi and Jiao, Wenpin},
  journal={ACM Transactions on Software Engineering and Methodology},
  volume={33},
  number={7},
  pages={1--30},
  year={2024},
  publisher={ACM New York, NY}
}

@article{li2025structured,
  title={Structured chain-of-thought prompting for code generation},
  author={Li, Jia and Li, Ge and Li, Yongmin and Jin, Zhi},
  journal={ACM Transactions on Software Engineering and Methodology},
  volume={34},
  number={2},
  pages={1--23},
  year={2025},
  publisher={ACM New York, NY}
}

\end{document}